\documentclass[english,12pt]{article}
\usepackage{array}
\usepackage{graphicx}
\usepackage{amssymb}
\usepackage{amsmath}
\usepackage{multirow}
\usepackage{prettyref}
\usepackage{babel}
\usepackage{units}
\usepackage[latin1]{inputenc}
\usepackage{amsfonts}
\usepackage{amssymb}
\usepackage{babel}
\usepackage{cite}
\def\@fmsl@sh#1#2#3{\m@th\ooalign{$\hfil#1\mkern#2/\hfil$\crcr$#1#3$}}
 \def\eq#1\en{\begin{equation}#1\end{equation}}
\def\s[#1,#2]{[#1\stackrel{\star}{,}#2]}
\def\sx[#1,#2]{[#1\stackrel{\star_{x}}{,}#2]}

\newcommand{\nc}{\newcommand}
\nc{\beq}{\begin{equation}}
\nc{\eeq}{\end{equation}}
\nc{\beqa}{\begin{eqnarray}}
\nc{\eeqa}{\end{eqnarray}}

\def\bc{\begin{center}}
\def\ec{\end{center}}

\def\gsim{\mathrel{\mathpalette\atversim>}}

\def\bc{\begin{center}}
\def\ec{\end{center}}

\def\gsim{\mathrel{\rlap{\lower4pt\hbox{\hskip1pt$\sim$}}

    \raise1pt\hbox{$>$}}}       %greater than or approx. symbol

\def\gsim{\mathrel{\rlap{\lower4pt\hbox{\hskip1pt$\sim$}}
    \raise1pt\hbox{$>$}}}       %greater than or approx. symbol

\begin{document}
\makeatletter
\def\fmslash{\@ifnextchar[{\fmsl@sh}{\fmsl@sh[0mu]}}
\def\fmsl@sh[#1]#2{%
  \mathchoice
    {\@fmsl@sh\displaystyle{#1}{#2}}%
    {\@fmsl@sh\textstyle{#1}{#2}}%
    {\@fmsl@sh\scriptstyle{#1}{#2}}%
    {\@fmsl@sh\scriptscriptstyle{#1}{#2}}}
\def\@fmsl@sh#1#2#3{\m@th\ooalign{$\hfil#1\mkern#2/\hfil$\crcr$#1#3$}}
\makeatother
%\baselineskip 24pt

%%%%%%%%%%%%%%%%%%%%%%%%%%%%%%%%%%%%%%%%%%%%%%%%%%%%%%%%%%%%%%%%%
%%%
%%%                      TITLE PAGE
%%%
%%%%%%%%%%%%%%%%%%%%%%%%%%%%%%%%%%%%%%%%%%%%%%%%%%%%%%%%%%%%%%%%%
\thispagestyle{empty}
\begin{titlepage}
\boldmath
\begin{center}
  \Large {\bf Higgs Starobinsky Inflation}
    \end{center}
\unboldmath
\vspace{0.2cm}
\begin{center}
{  {\large Xavier Calmet}\footnote{x.calmet@sussex.ac.uk}
and {\large Iber\^ e Kuntz}\footnote{ibere.kuntz@sussex.ac.uk}}
 \end{center}
\begin{center}
{\sl Physics $\&$ Astronomy, 
University of Sussex,   Falmer, Brighton, BN1 9QH, United Kingdom 
}
\end{center}
\vspace{5cm}
\begin{abstract}
\noindent
In this paper we point out that Starobinky inflation could be induced by quantum effects due to a large non-minimal  coupling of the Higgs boson to the Ricci scalar. The Higgs Starobinsky mechanism provides a solution to issues attached to large Higgs field values in the early universe which in a metastable universe would not be a viable option. We verify explicitly that these large quantum corrections do not destabilize Starobinsky's potential. 
\end{abstract}  
\end{titlepage}

%\pacs{}

%%%%%%%%%%%%%%%%%%%%%%%%%%%%%%%%%%%%%%%%%%%%%%%%%%%%%%%%%%%%%%%%
%%%
%%%                     INTRODUCTION
%%%
%%%%%%%%%%%%%%%%%%%%%%%%%%%%%%%%%%%%%%%%%%%%%%%%%%%%%%%%%%%%%%%%

\newpage

The idea that inflation may be due to degrees of freedom already present in the standard model of particle physics or quantum general relativity is extremely attractive and has received much attention in the recent years. In particular two models stand out by their simplicity and elegance. Higgs inflation  \cite{Bezrukov:2007ep,Barvinsky:2008ia,Barvinsky:2009fy} with a large non-minimal coupling of the Higgs boson $H$ to the Ricci scalar ($\xi H^\dagger H R$) and Starobinsky's inflation model \cite{Starobinsky:1980te} based on $R^2$ gravity are both minimalistic and perfectly compatible with the latest Planck data.

These two models should not be considered as physics beyond the standard model but rather both operators $\xi H^\dagger H R$  and $R^2$ are expected to be generated when general relativity is coupled to the standard model of particle physics. We will come back to that point shortly. The aim of this paper is to point out an intriguing distinct possibility, namely that Starobinsky inflation is generated by quantum effects due to a large non-minimal coupling of the Higgs boson to the Ricci scalar. In that framework, we do not need to posit that the Higgs boson starts at a high field value in the early universe which would alleviate constraints coming from the requirement of having a stable Higgs potential even for large Higgs field values \cite{Kobakhidze:2014xda,Degrassi:2012ry,Bezrukov:2012sa}.

We shall now argue that both terms necessary for Higgs inflation or Starobinsky's model are naturally present when the standard model of particle physics is coupled to general relativity. While the quantization of general relativity remains one of the outstanding challenges of theoretical physics, it is possible to use effective field theory methods below the energy scale $M_\star$ at which quantum gravitational effects are expected to become large. The energy scale $M_\star$ is usually assumed to be of the order of the Planck scale $M_P=\sqrt{8 \pi G_N}^{-1}=2.4335 \times 10^{18}$ GeV, however recent work has shown that even in four space-time dimensions this energy scale is model dependent. At energies below $M_\star$, we can describe all of particle physics and cosmology with the following effective field theory  (see e.g. \cite{Codello:2015mba,Donoghue:2014yha,BD})
\begin{eqnarray}\label{action1} 
S &=& \int d^4x \, \sqrt{-g} \biggl( \left( \frac{1}{2}  M^2 + \xi H^\dagger H \right)  R- \Lambda_C^4  + c_1 R^2 + c_2 C^2 +c_3 E + c_4 \Box R+  \\  \nonumber &&  -  L_{SM} - L_{DM} + O(M_\star^{-2})   \biggr ) 
\end{eqnarray}
where we have restricted our considerations to dimension four operators which are expected to dominate at least at low energies. Note that we are using the Weyl basis and the following notations: $R$ stands for the Ricci scalar, $R^{\mu\nu}$ for the Ricci tensor,  $E=R_{\mu\nu\rho\sigma}R^{\mu\nu\rho\sigma}- 4 R_{\mu\nu}R^{\mu\nu}+R^2$, $C^2=E+2R_{\mu\nu}R^{\mu\nu} -2/3 R^2$, the dimensionless $\xi$ is the non-minimal coupling of the Higgs boson $H$ to the Ricci scalar, the coefficients $c_i$  are dimensionless free parameters,  the cosmological constant $\Lambda_C$ is of order of $10^{-3}$ eV, the Higgs boson vacuum expectation value, $v=246$ GeV  contributes to the value of the Planck scale 
\begin{eqnarray}
\label{effPlanck}(M^2+\xi v^2)=M_P^2 \, ,
\end{eqnarray}
$L_{SM}$ contains all the usual standard model interactions (including mass terms for neutrinos) and finally $L_{DM}$ describes the dark matter sector (this is the only part of the model which has not been tested yet experimentally). Submillimeter  pendulum tests of Newton's law \cite{Hoyle:2004cw} lead to extremely weak limits on the parameters  $c_i$. In the absence of accidental cancellations between these coefficients, they are  constrained to be less than $10^{61}$ \cite{Calmet:2008tn}. The discovery of the Higgs boson and precision measurements of its couplings to fermions and bosons at the LHC can be used to set a limit on $\xi$.  One finds that $|\xi| < 2.6 \times 10^{15}$  \cite{Atkins:2012yn}. Clearly very little is known about the values of $c_i$ and $\xi$.

Besides describing all of particle physics and late time cosmology, the action given in Eq. (\ref{action1}) can also describe inflation if some of its parameters take specific values and if some of its fields fulfil specific initial conditions in the early universe. This action, depending on the initial conditions can describe either Higgs inflation if $\xi \sim 10^4$  and the Higgs field is chosen to take large values in the early universe or Starobinsky inflation if $c_1 \sim 10^9$ and the corresponding scalar extra degree of freedom which can be made more visible by going to the Einstein frame takes large values in the early universe.

If we assume that the Higgs fields take small values in the early universe,  Eq. (\ref{action1}) reduces to 
\begin{eqnarray}
S_{Starobinsky}^J=\int d^4 x \sqrt{g} \frac{1}{2} \left  (M_P^2 R+ c_S R^2   \right )
\end{eqnarray}
during inflation  which in the Einstein frame gives
\begin{eqnarray}
S_{Starobinsky}^E=\int d^4 x \sqrt{g} \left    (\frac{M_P^2}{2} R - \frac{1}{2} \partial_\mu \sigma \partial^\mu \sigma - \frac{M_P^4}{c_S} \left (1-\exp \left( - \sqrt{\frac{2}{3}} \frac{\sigma}{M_P}\right) \right)^2  \right ).
\end{eqnarray}
We have assumed that the scalar degree  of freedom $\sigma$ hidden in $R^2$ takes large field values in the early universe. A successful prediction of the density perturbation $\delta \rho/\rho$ requires $c_S=0.97 \times 10^9$ \cite{Netto:2015cba,Staro}. On the other hand,  if we assume that only the Higgs field takes large values in the early universe,the action (\ref{action1}) reduces to
\begin{eqnarray}
S_{Higgs}^J&=&\int d^4 x \sqrt{-g} \left  (\frac{M^2}{2} R+ \xi_H H^\dagger H R  - L_{SM} \right ) \\ &=& \int d^4 x \sqrt{-g} \left  (\frac{M^2+\xi_H h^2}{2} R -\frac{1}{2} \partial_\mu h \partial^\mu h +\frac{\lambda}{4} (h^2-v^2)^2 \right)+\ldots \nonumber.
\end{eqnarray}
In the Einstein frame, one obtains
\begin{eqnarray}
S_{Higgs}^E=\int d^4 x \sqrt{\hat g} \left    (\frac{M_P^2}{2} \hat R -\frac{1}{2} \partial_\mu \chi \partial^\mu \chi +U(\chi) +\ldots \right ) 
 \end{eqnarray}
with
\begin{eqnarray}
\frac{d\chi}{d h}=\sqrt{\frac{\Omega^2 + 6 \xi_H^2 h^2/M_P^2}{\Omega^4}}
\end{eqnarray}
where $\Omega^2=1+ \xi_H^2 h^2/M_P^2$
and
\begin{eqnarray}
U(\chi)=\frac{1}{\Omega(\chi)^4} \frac{\lambda}{4} (h(\chi)^2 -v^2)^2.
\end{eqnarray}
A successful prediction of the density perturbation $\delta \rho/\rho$ requires $\xi_H=1.8 \times 10^4$.

These two models are very attractive because they do not necessitate physics beyond the standard model. Furthermore, they are compatible with current cosmological observations which favor small tensor perturbations that so far have not been observed.  It has actually been pointed out that both models are phenomenologically very similar \cite{Bezrukov:2011gp,Salvio:2015kka}. However, while Starobinky's inflation model does not suffer from any obvious problem, it has recently been pointed out that in the case of Higgs inflation, which necessitate the Higgs field to take very large field values, our universe will not end up in the standard model Higgs vacuum if it is metastable as suggested by the latest measurement of the top quark mass, but rather in the real vacuum of the theory which does not correspond to the world we observe. In this paper we point out that there is an alternative possibility. Namely when quantum corrections are taken into account, a large non-minimal coupling of the Higgs boson can generate Starobinsky inflation by generating a large coefficient for the coefficient of $R^2$ in the early universe.  While the model corresponds to Starobinsky's model, the Higgs boson plays a fundamental role as it triggers inflation by generating a large coefficient for $R^2$.

The action given in Eq. (\ref{action1}) needs to be renormalized. We will work in dimensional regularization  to avoid having to discuss the dependence of observables on the cutoff (this problem is due to the non-renormalizability of quantum gravity). We shall neglect the cosmological constant which is not important for inflation purposes. In that case, Newton's constant does not receive any correction to leading order. On the other hand, the coefficient  $c_1$ of $R^2$ gets renormalized and one can define a renormalization group equation for this coupling constant.  $N_s$ scalar fields with a non-minimal coupling to the Ricci scalar $\xi$  will lead to the following renormalization group equation \cite{Codello:2015mba,Donoghue:2014yha,BD}
\begin{eqnarray}
\mu \partial_\mu c_1(\mu)=\frac{(1- 12 \xi)^2}{1152 \pi^2} N_s
\end{eqnarray}
to leading order (i.e. neglecting the graviton contribution which is suppressed by $1/\xi$), note that fermions and vector fields do not contribute to the renormalization of $c_1$ in the Weyl basis. The renormalization group equation can be easily integrated, one finds \cite{Codello:2015mba,Donoghue:2014yha,BD}:
\begin{eqnarray}
c_1(\mu_2)=c_1(\mu_1)+\frac{(1- 12 \xi)^2 N_s}{1152 \pi^2} \log \frac{\mu_2}{\mu_1}.
\end{eqnarray}
The bounds on $c_1$ in today's universe are very weak as mentioned before. Even if $c_1(today)$ is of order unity, it would have been large in the early universe if the Higgs non-minimal coupling $\xi$ is large. Indeed, we assume that inflation took place at some high energy scale e.g. $\mu \sim10^{15}$ GeV, the $\log$ term is a factor of order 60 if we take the scale $\mu_1$ of the order of the cosmological constant. A Higgs non-minimal coupling to the Ricci scalar of $\xi=1.8 \times 10^4$ would lead to a coefficient $c_1=0.97\times 10^9$ for $R^2$. Assuming that the scalar extra degree contained in $R^2$ took large field values in the early universe, a large non-minimal coupling of the Higgs boson to the Ricci scalar can trigger Starobinsky inflation even if the standard model vacuum is metastable as the Higgs boson itself does not roll down its potential during inflation. Inflation is due entirely to the $R^2$, but is triggered by the Higgs large non-minimal coupling. 

Let us emphasize two important points. The first one is that $c_1\sim 0.97\times 10^9$ is fixed by the CMB constraint. This parameter only takes such a large value at inflationary energy scales due to its renormalization group evolution. The second one, is that we are neglecting the running of the Higgs boson non-minimal coupling to the Ricci scalar. However, this is a very good approximation. The leading contributions of the standard model to the beta-function of the non-minimal coupling are known \cite{Buchbinder} :
\begin{eqnarray}
\beta_\xi =\frac{6 \xi+1}{(4 \pi)^2} \left [ 2 \lambda + y_t^2 -\frac{3}{2} g^2 -\frac{1}{4} g^{\prime 2} \right ]
\end{eqnarray}
where $\lambda$ is the self-interaction coupling of the Higgs boson, $g$ the SU(2) gauge coupling and $g^\prime$ the U(1) gauge coupling. Quantum gravitational corrections will be suppressed by powers of the Planck mass and can thus be safely ignored as long as we are at energies below the Planck mass.

One might worry that if the large non-minimal coupling of the Higgs boson triggers a large coefficient for the operator $R^2$, it might also generate new terms in the effective action which could destabilize the potential.  The leading order effective action to the second order in the curvature expansion induced by scalar fields non-minimally coupled to gravity is known \cite{Codello:2015mba,Donoghue:2014yha}:
\begin{eqnarray} \label{EFT}
S_{EFT}&=& \frac{1}{16 \pi G} \int d^4x \sqrt{-g} \left ( R + \alpha R^2 + \beta R \log \frac{-\Box}{\mu^2} R +  \gamma  C^2 + \ldots\right).
\end{eqnarray}
Note that here we are neglecting the cosmological constant, $\alpha= c_1 \times16 \pi G $ and $\gamma = c_2 \times16 \pi G $ are renormalized coupling constants and we shall assume that $c_2$ is small at the scale of inflation, it is not sensitive to the Higgs boson's non-minimal coupling, while we have fixed the Higgs non-minimal coupling such that $c_1=0.97\times 10^9 $.  The coefficient $\beta$ is a prediction of the effective action and is given by $N_s (1-12 \xi)^2/(2304 \pi^2)  \times 16 \pi G$ where $N_s$ is the number of scalar field degrees of freedom in the model, in our case 4. The coefficient $N_s (1- 12\xi)^2/(2304 \pi^2)$  is indeed large  and of the order of $7.8 \times 10^6$ and we have to check that the log-term does not lead to sizable contributions to the effective potential of the Starobinsky's field. Before verifying this explicitly, let us mention that the large non-minimal coupling between the Higgs boson and the Ricci scalar which is necessary to induce Starobinsky inflation does not lead to perturbative unitarity problems \cite{Calmet:2013hia} (see Appendix A).

Note that the coefficients of $E$ and of $C^2$ do not depend on the non-minimal coupling of the Higgs boson to the Ricci scalar. Furthermore in 4 dimensions, $E$ does not contribute to the equations of motion. The coefficient of the term $C^2$ is assumed before renormalization to be of the same order as that of $R^2$, i.e. of order 1. However, after renormalization the coefficient of $R^2$ is tuned to be very large and of the order of $10^9$ while the coefficient of $C^2$ remains small compared to the renormalized coefficient of $R^2$. $C^2$ is thus negligible. 

We shall treat the effective action (\ref{EFT}) as a $F(R)$ gravity with $F(R)=R + \alpha R^2 + \beta R \log \frac{-\Box}{\mu^2} R$. There is a well established procedure to map such models from the Jordan frame to the Einstein frame, see e.g. \cite{Sebastiani:2015kfa}. The potential for the inflaton in the Einstein frame is given by 
\begin{eqnarray} \label{potential}
V(\phi)= \frac{1}{2 \kappa^2} \left ( e^{\sqrt{\frac{2}{3}} \kappa \phi} R(\phi) - e^{2 \sqrt{\frac{2}{3}} \kappa \phi} F(R(\phi)) \right )
\end{eqnarray}
where $\kappa^2 = 8 \pi G$ and $R(\phi)$ is a solution to the equation
\begin{eqnarray}
\phi = -\sqrt{\frac{3}{2}} \frac{1}{\kappa} \log \frac{dF(R)}{dR}.
\end{eqnarray}
We can find a formal solution to this equation
\begin{eqnarray}
R(\phi)= \frac{1}{2 \alpha } \left ( \frac{1}{1+\frac{\beta}{2 \alpha} \log \left ( \frac{-\Box}{\mu^2} \right)} \right ) \left ( e^{-\sqrt{\frac{2}{3}} \kappa \phi} -1 \right ).
\end{eqnarray}
This expression for $R(\phi)$ can be understood as a series $\frac{\beta}{2 \alpha}$ which is a small parameter:
\begin{eqnarray}
R(\phi)= \frac{1}{2 \alpha } \left (1 - \sum_{n=1}^\infty (-1)^{n+1} \left (  \frac{\beta}{2 \alpha} \log \left ( \frac{-\Box}{\mu^2} \right)\right )^n \right ) \left ( e^{-\sqrt{\frac{2}{3}} \kappa \phi} -1 \right ).
\end{eqnarray}
where the $ \log$-term can be expressed using
\begin{eqnarray}
\log \left ( \frac{-\Box}{\mu^2} \right) = \int_0^\infty ds \left ( \frac{1}{\mu^2 + s} - \frac{1}{-\Box + s} \right ).
\end{eqnarray}
The zeroth order term in $\frac{\beta}{2 \alpha}\sim 4 \times 10^{-3}$ corresponds to the usual Starobinsky solution:
\begin{eqnarray}
R(\phi)^{(0)}=R(\phi)_{Starobinsky}= \frac{1}{2 \alpha }  \left ( e^{-\sqrt{\frac{2}{3}} \kappa \phi} -1 \right ).
\end{eqnarray}
The series expansion will generate  higher order terms corresponding to operators of the type 
$\exp({-\sqrt{\frac{2}{3}} \kappa \phi}) (2/3 \kappa^2 \partial_\mu \phi \partial^\mu \phi -\sqrt{2/3} \kappa \Box \phi)$ and higher derivatives thereof. These new terms are however suppressed by powers of $\frac{\beta}{2 \alpha}$ and can be safely ignored. It is easy to check that the $\log$-term appearing in the $F(R)$ term of the potential (\ref{potential}) is also suppressed by $\frac{\beta}{2 \alpha}$ compared to the usual Starobinsky's potential.  

 We conclude that the large quantum corrections induced by the large Higgs boson non-minimal coupling do not affect the flatness of Starobinsky's potential. Let us add a few remarks. The model discussed above is not a new model. Physics (including reheating or preheating and all of particle physics) is identical to that predict in Starobinsky's model. We merely identify a new connection between the Higgs boson and inflation. As in the case of the standard Starobinsky model, a coupling $\phi^2 h^2$ will be generated. It is however suppressed by factors of  $m_{Higgs}^2/M_P^2$ which is a small number, particle physics will thus not be affected and the Higgs boson behaves as the standard model Higgs boson. Furthermore, the Higgs field does not take large values in the early universe, we can thus safely ignore the term $H^\dagger H R$ when studying the inflationary potential. Note that there are subtleties when considering the equivalence of quantum corrections in different parameterizations/representations of the theory (i.e. when going from the Jordan frame to the Einstein frame). Here we are avoiding this problem: we renormalized the theory in the Jordan frame where the model is defined and then map the effective action to the Einstein frame. When proceeding this way, there are no ambiguities or risk to mix up the orders in perturbation theory and the expansion in the conformal factor (see e.g. \cite{Calmet:2012eq,Kamenshchik:2014waa,Vilkovisky:1984st}) .

In this paper, we have identified a new connection between the Higgs boson and inflation. In the model envisaged here, the Higgs boson is not the inflaton but it generates inflation by creating a large Wilson coefficient for the $R^2$ operator and it is thus at the origin of Starobinsky's inflation. This mechanism is interesting as it does not require physics beyond the standard model. The Higgs boson does not need to take large field values in the early universe and we could thus be living in a metastable potential. 

{\it Acknowledgments:}
This work is supported in part by the Science and Technology Facilities Council (grant number  ST/L000504/1) and by the National Council for Scientific and Technological Development (CNPq - Brazil). 

\section*{Appendix A}
 It has been shown in \cite{Calmet:2013hia} that a large non-minimal coupling of the Higgs to the Ricci scalar does not lead to a new physical scale. While perturbative unitarity appears to be naively violated at an energy scale of $M_P/\xi$, it can be shown by resumming an infinite series of one-loop diagrams in the large $\xi$ and large $N$ limits but keeping $\xi G_N N$ small that perturbative unitarity is restored (this phenomenon has been called self-healing by Donoghue).  In this limit one finds 
 \begin{eqnarray}
 i D^{\alpha \beta \mu \nu}_{dressed}= -\frac{i}{2 s} \frac{L^{\alpha \beta} L^{\mu \nu}}{\left (1- \frac{ s F_1(s)}{2}\right) }.
 \end{eqnarray} 
where $L^{\alpha \beta}=\eta^{\alpha\beta}-q^\alpha q^\beta/q^2$ and
\begin{eqnarray}
F_1(q^2)=-\frac{1}{30 \pi} N_s G_N(\bar h)(1+10\xi + 30 \xi^2) \log\left( \frac{-q^2}{\mu^2}\right).
\end{eqnarray} 
The background dependent Newton's constant is given by
\begin{eqnarray}
G_N(\bar h)
=
\frac{1}{8 \pi (M^2+\xi \bar h^2)}
\ .
\end{eqnarray} 
In the model described in this paper, one has $\bar h=v$. Note that $F_1(s)$ is negative, there is thus no physical pole in the propagator.  The dressed amplitude in the large $\xi$ and large $N$ limits is given by
  \begin{eqnarray}
  A_{dressed}=\frac{48 \pi G_N(\bar h) s  \xi^2}{1+\frac{2}{\pi} G_N(\bar h)  s \xi^2 \log(-s/\mu^2)} 
    \end{eqnarray}
One easily verifies that the $J=0$ partial-wave dressed amplitude fulfils 
\begin{eqnarray}
|a_{0}|^2=\mbox{Im} \left( a_{0}\right).
\end{eqnarray} 
In other words, unitarity is restored within general relativity without any new physics or strong dynamics (we are keeping $\xi G_N$ small) and there is no new scale associated with the non-minimal coupling despite naive expectations. The cut-off of the effective theory is thus the usual Planck scale.

%%%%%%%%%%%%%%%%%%%%%%%%%%%%%%%%%%%%%%%%%%%%%%%%%%%%%%%%%%%%%%%%%
%%%
%%%                     BIBLIOGRAPHY
%%%
%%%%%%%%%%%%%%%%%%%%%%%%%%%%%%%%%%%%%%%%%%%%%%%%%%%%%%%%%%%%%%%%%

\bigskip{}

\end{document}